\documentclass[pra,twocolumn]{revtex4}
\usepackage{graphicx}
\usepackage{amsfonts,amsmath,amssymb,float}
\usepackage{bm,dsfont}
\usepackage{color}
\usepackage{bbm}
\usepackage{braket}
\usepackage{longtable}
\usepackage[dvips]{epsfig}
\usepackage{amsmath,amssymb,lscape,float}
\usepackage{hyperref}
\usepackage{listings}

\begin{document}
\title{Spectral features of polaronic excitations in a superconducting analog simulator}

\author{Julian K. Nauth and Vladimir M. Stojanovi\'c}
\affiliation{Institut f\"{u}r Angewandte Physik, Technical University of Darmstadt, 
D-64289 Darmstadt, Germany}

\date{\today}

\begin{abstract}
We investigate spectral properties of polaronic excitations within the framework of an analog quantum simulator 
based on inductively coupled superconducting transmon qubits and microwave resonators. This system emulates a 
lattice model that describes a nonlocal coupling of an itinerant spinless-fermion excitation to dispersionless 
(Einstein-type) phonons through the Peierls- and breathing-mode interaction mechanisms. The model is characterized 
by a sharp, level-crossing transition at a critical value of the effective excitation-phonon coupling strength; 
above the transition point, the ground state of this model corresponds to a heavily dressed (small-polaron) excitation. 
Using the kernel-polynomial method, we evaluate the momentum-frequency resolved spectral function of this system 
for a broad range of parameters. In particular, we underscore the ramifications of the fact that the zero-quasimomentum 
Bloch state of a bare excitation represents the exact eigenstate of the Hamiltonian of this system for an arbitrary 
excitation-phonon coupling strength. We also show that -- based on the numerically evaluated spectral function and 
its well-known relation with the survival probability of the initial, bare-excitation Bloch state (the Loschmidt 
echo) -- one can make predictions about the system dynamics following an excitation-phonon interaction quench. 
To make contact with anticipated experimental realizations, we utilize a previously proposed method for extracting 
dynamical-response functions in systems with local (single-qubit) addressability using the multi-qubit (many-body)
version of the Ramsey interference protocol.
\end{abstract}
\maketitle

\section{Introduction}
The field of quantum simulation~\cite{Georgescu+:14} rose to prominence over the past decade as a thriving research area 
at the intersection of the traditional fields of condensed-matter- and atomic, molecular, and optical physics, on the one 
hand, and the rapidly-developing field of quantum information processing on the other. In particular, analog quantum simulators 
provide valuable insights into various properties of complex many-body systems~\cite{Hohenadler+:12,Gangat+:13,Kapit:13,
Yang+:18,LasHeras++:14,Egger+Wilhelm:13,Leppakangas+:18}, while at the same time typically requiring much more modest quantum-hardware 
resources -- i.e., a smaller number of qubits -- for their realization than what will eventually be required for large-scale 
universal quantum computation. Examples of physical platforms usually employed for implementing analog quantum simulators 
are furnished by those based on cold neutral atoms in optical lattices~\cite{Bloch+Dalibard+Zwerger:08} or tweezers~\cite{MorgadoWhitlock:21}, 
trapped ions\cite{Bruzewicz+:19}, cold polar molecules~\cite{Gadway+Yan:16}, and superconducting (SC) quantum circuits~\cite{Paraoanu:14,
Wendin:17}. The latter usually entail transmon qubits~\cite{Koch+:07} and SC microwave resonators, the key building blocks 
of circuit quantum electrodynamics (circuit-QED) systems~\cite{Wallraff+:04,Girvin:14,Blais+:21}; analog simulators with similar 
functionalities can also be realized with flux qubits~\cite{Volkov+Fistul:14,Fistul+:22}.

Analog simulators of quantum many-body systems are typically designed with the aim of investigating their ground-state- (static) 
or dynamical properties~\cite{Georgescu+:14}. On the other hand, it was only recently that some attention was also dedicated to 
elucidating {\em spectral properties} of such systems by virtue of analog simulators~\cite{Hendry+Feiguin:19,Baez++:20}. Motivated
in part by the apparent dearth of studies pertaining to this important aspect of simulated many-body systems, the present work 
is devoted to spectral properties of a one-dimensional system that comprises an itinerant spinless-fermion excitation coupled to 
dispersionless (zero-dimensional) phonons through two nonlocal coupling mechanisms -- more precisely, Peierls- and breathing-mode 
type excitation-phonon (e-ph) interactions~\cite{Stojanovic+:14,Stojanovic+Salom:19}. The present study is framed within the context 
of an analog SC quantum simulator, which was previously proposed for studying both static~\cite{Stojanovic+:14} and 
dynamical~\cite{Stojanovic+Salom:19} properties of this coupled e-ph system. The principal building blocks of this 
simulator are transmon qubits and microwave resonators~\cite{VoolDevoretReview:17,Wendin:17,Gu+:17,SCdevicesPractical:21,
SCcircuitsTutorial:21}. Importantly, the qubit-resonator coupling in this simulator belongs to the class of indirect inductive 
couplings~\cite{SCqubitReview:19}, and the physical mechanism that enables it entails a flux of the resonator (microwave-photon) 
modes that pierces specially tailored coupler circuits connecting adjacent qubits. In this system, SC qubits are coupled through 
nearest-neighbor $XY$-type (flip-flop) interaction whose effective strength depends dynamically on the resonator degrees of freedom
(i.e. photons), this dependence being equivalent to that of the $XY$ spin-Peierls model~\cite{Cross:79,Barford:05}.

The quantity of primary interest in our present context is the {\em momentum-frequency resolved spectral function}, this last dynamical 
response function being closely related to the Fourier transform of the single-particle retarded Green's function~\cite{FoersterBOOK}. 
This spectral function, which captures {\em inter alia} the essential features of dressed polaronic excitations in the strong-coupling 
regime of the system under consideration, will be evaluated here in a numerically-exact manner. To be more specific, the {\em kernel-polynomial method} (KPM)~\cite{WeisseEtAlRMP:06} -- based on the expansion of the relevant spectral function in Chebyshev polynomials
of the first kind -- will be utilized here for computing the spectral function for various choices of parameters characterizing 
the SC analog simulator under consideration. The KPM, pioneered by Silver and R\"{oder}~\cite{Silver+Roeder:97}, was successfully 
employed in the past for evaluating both zero-~\cite{Alvermann+Fehske:08} and finite-temperature~\cite{Schubert+:05} dynamical response functions in a variety of quantum many-body systems~\cite{WeisseEtAlRMP:06}. Motivated by the need to accurately compute spectral 
densities of strongly-interacting quantum many-body systems, various generalizations of this method have also been 
proposed~\cite{Sobczyk+Roggero:22}.

What makes the proposed analysis of momentum-frequency resolved spectral functions within the framework of a SC analog simulator 
particularly pertinent is the availability of a method for the experimental measurements of such quantities and their counterparts 
in the time domain~\cite{Knap++:13}. This method is based on a generalized, multi-qubit version of the Ramsey interference protocol~\cite{Ramsey:50} 
(recall that, when applied to a single SC qubit, the Ramsey interference protocol is used to determine its dephasing time 
$T_2$~\cite{Rigetti++:12}) and is applicable to all locally-addressable systems~\cite{Stojanovic+:14}. In particular, being amenable 
to experimental control at the single-qubit level, the envisioned SC analog simulator constitutes a nearly ideal platform for 
the practical demonstration of this method.

Heavily-dressed quasiparticles formed in the strong-coupling regime of systems that feature short-range interaction of an itinerant 
excitation (electron, hole, exciton) with dispersionless (Einstein-like) phonons are referred to as {\em small polarons}. While the 
Holstein model~\cite{Holstein:59}, describing local e-ph interaction~\cite{Jeckelmann+White:98,Bonca+:99,Ku+Bonca:02}, is the most 
common starting point for investigating such quasiparticles, more realistic models that involve nonlocal e-ph coupling mechanisms 
have come to the fore
over the last decade. These mechanisms -- exemplified by Peierls-~\cite{Stojanovic+:04} and breathing-mode type~\cite{Slezak++:06} e-ph 
couplings -- are known to play important roles in certain classes of narrow-band electronic materials~\cite{Hannewald:04,Hannewald++:04,
Roesch+:05,Vukmirovic+:10,Stojanovic++:10,Ciuchi+Fratini:11,Vukmirovic+:12,Makarov+:15,Shneyder+:20,Shneyder+:21}. Moreover, models involving 
Peierls-type coupling show sharp, level-crossing transitions in the ground-state-related quantities at certain critical values of the 
effective e-ph coupling strength~\cite{Stojanovic+Vanevic:08}; such couplings, whose corresponding e-ph vertex functions depend on both
excitation- and phonon quasimomenta, do not belong to the realm of validity of the time-honoured Gerlach-L\"{o}wen theorem~\cite{Gerlach+Lowen:87} that rules out such nonanalyticities for most known e-ph interaction types. Finally, those models also display nontrivial 
e-ph entanglement properties, both in the ground-~\cite{Stojanovic+Vanevic:08,Stojanovic:20} and excited states~\cite{Roosz+Held:22}. 

Several analog simulators of the Holstein model have as yet been proposed that make use of physical platforms as diverse as trapped
ions~\cite{Stojanovic+:12}, cold polar molecules~\cite{HerreraEtAl}, and SC circuits~\cite{Mei+:13}. On the other hand, models that 
involve nonlocal (momentum-dependent) e-ph interactions have as yet only been simulated using arrays of SC qubits and microwave
resonators~\cite{Stojanovic+:14,Stojanovic+Salom:19}, where photons in resonators play the role of phonons. The previous studies 
covered both ground-state-~\cite{Stojanovic+:14} and dynamical~\cite{Stojanovic+Salom:19} properties of small polarons in such models, 
which motivates us to also investigate their spectral properties in the present work.

The outline of the remainder of this paper is as follows. In Sec.~\ref{genspectfunc} spectral poperties of 
dressed excitations are briefly reviewed, starting with general features of momentum-frequency resolved 
spectral functions. This is followed by a brief summary of the Ramsey-protocol based method for extracting 
spectral functions experimentally in Sec.~\ref{Extract_via_Ramsey}. In Sec.~\ref{SimulatorAndHamiltonian}, 
the analog simulator under consideration and its governing Hamiltonian are introduced, along with a brief 
descripton of the derivation of its effective Hamiltonian. The numerical results obtained for the momentum-frequency 
resolved spectral function of this system are presented and discussed in Sec.~\ref{resdiscuss}. Finally, 
the paper is summarized with some general concluding remarks in Sec.~\ref{sumconcl}. Some mathematical details 
pertaining to the form of the qubit-qubit interaction in the system at hand, the truncation of the Hilbert 
space of the system, and the use of the KPM for computing the spectral function, are relegated to 
Appendices~\ref{DerivCosform}, \ref{ExactDiag}, and \ref{KPMforSpectFunc}, respectively.

\section{Momentum-frequency resolved spectral function}   \label{genspectfunc}
Dynamical response functions, defined through Fourier transforms of retarded two-time correlation
functions, provide a framework for characterizing excitations in many-body systems and are intimately
related to their resulting magnetic, optical, and transport properties~\cite{FoersterBOOK}. Among the 
relevant two-time correlation functons, a particularly important example is furnished by the single-particle
retarded Green's function, which conventionally describes the propagation of a single electron (or a hole). 
Here this Green's function will be employed in the context of an itinerant spinless-fermion excitation that 
interacts with zero-dimensional bosons residing on sites of a one-dimensional lattice.

The single-particle retarded Green's function of interest in the problem under consideration is given by
\begin{equation}\label{anticommGF}
G_{+}^{\textrm{R}}(k,t)=-\frac{i}{\hbar}\:\theta(t)\langle
\textrm{G}_0|[c_k^{\dagger}(t),c_k]_{+}|\textrm{G}_0\rangle \:.
\end{equation}
Here $c_k^{\dagger}(t)\equiv U^{\dagger}_{H}(t)c_k^{\dagger}U_{H}(t)$ is a single-particle creation 
operator in the Heisenberg representation, where $U_{H}(t)$ is the time-evolution operator of the system, 
with $H$ being its governing Hamiltonian; $\theta(t)$ is the Heaviside step function, $|\textrm{G}_0\rangle$ 
the ground state of the system, and $[\ldots]_{+}$ denotes an anticommutator. 

The last retarded Green's function describes the linear response of the system to the addition and removal of 
a single fermion. To formally evaluate its Fourier transform, a regularization factor $\lim_{\eta\rightarrow
0^{+}}e^{-\eta|t|}$ ought to be included. Accordingly, the regularized Fourier transform of $G_{+}^{\textrm{R}}(k,t)$ 
is given by
\begin{equation}
G_{+}^{\textrm{R}}(k,\omega)=\int^{\infty}_{-\infty}G_{+}^{\textrm{R}}(k,t)
e^{i\omega t-\eta|t|}dt \quad (\:\eta\rightarrow 0^{+}\:)\:,
\end{equation}
and -- with the aid of Eq.~\eqref{anticommGF} -- leads to
\begin{eqnarray}
G_{+}^{\textrm{R}}(k,\omega)&=&\langle\textrm{G}_0|c_k\:
\frac{1}{\hbar\omega+i0^{+}+E_0-H}
\:c_k^{\dagger}|\textrm{G}_0\rangle \nonumber \\
&+& \langle\textrm{G}_0|c_k^{\dagger}\:\frac{1}
{\hbar\omega+i0^{+}-E_0+H}\:c_k|\textrm{G}_0\rangle\:,
\end{eqnarray}
where $E_0$ is the ground-state energy of the system. 

The quantity of primary interest for the present work, the {\em momentum-frequency resolved 
spectral function}, is given by
\begin{equation} \label{CpectFuncDef}
A(k,\omega)=-\frac{1}{\pi}\:\textrm{Im}\:G_{+}^{\textrm{R}}(k,\omega)\:.
\end{equation}
This single-particle spectral function can be expressed explicitly in terms of the eigenstates 
and eigenvalues of the Hamiltonian $H$, namely as 
\begin{equation} \label{MomFreqSpectFunc}
A(k,\omega)=\sum_j\:|\langle\psi^{(j)}_k|c^{\dagger}_k |0\rangle|
^{2}\delta\left(\omega-E_k^{(j)}/\hbar\right) \:,
\end{equation}
where $|\psi^{(j)}_k\rangle$ are the eigenstates and $E_k^{(j)}$ the corresponding eigenvalues
of the Hamiltonian $H$ with quasimomentum $k$. This spectral function satisfies the sum rule
\begin{equation} \label{SumRuleSpectFunc}
\int_{-\infty}^{\infty} A(k,\omega)\:d\omega=1 
\end{equation}
for each quasimomentum $k$.

In the coupled e-ph system to be addressed in what follows, the last spectral function 
is intimately related to the system dynamics following an e-ph interaction quench. Namely, assuming the
initial ($t=0$) states to have the form of bare-excitation Bloch states $c^{\dagger}_k|0\rangle$ at different 
quasimomenta $k$ (where $|0\rangle$ is the vacuum state of the system), the spectral function $A(k,\omega)$ 
is given by the Fourier transform of the matrix element $\langle\psi(t)|c^{\dagger}_k |0\rangle$ (where 
$|\psi(t)\rangle$ is the state of the system at time $t$) -- the amplitude for the system to remain 
in the initial (bare-excitation) state at time $t$. This quantity represents the special case of what 
is more generally referred to as the Loschmidt amplitude and its squared module (i.e., the survival 
probability of the initial state) 
\begin{equation} \label{LoschmidtEcho}
\mathcal{L}_k(t)\equiv|\langle\psi(t)|c^{\dagger}_k |0\rangle|^{2}
\end{equation} 
is known as the Loschmidt echo~\cite{Peres:85}. The latter constitutes the most widely used quantity 
for characterizing nonequilibrium quantum dynamics~\cite{Heyl:18}.

\section{Measurement of retarded Green's function using Ramsey protocol} \label{Extract_via_Ramsey}
In the following, we first introduce the retarded single-particle Green's 
functions of relevance for the present work (Sec.~\ref{GF_subsec}). We then recapitulate 
the basics of the method for the experimental measurement of those Green's functions 
using many-body (multi-qubit) version of the Ramsey interference protocol (Sec.~\ref{Ramsey_subsec}),
which was proposed in Ref.~\cite{Knap++:13} and adapted for application in systems
of the kind discussed here in Ref.~\cite{Stojanovic+:14}.

\subsection{Retarded single-particle Green's functions} \label{GF_subsec}
While the anticommutator Green's function [cf. Eq.~\eqref{anticommGF}] is the appropriate 
one to use for spinless-fermion excitations in the problem under consideration, its commutator 
counterpart
\begin{equation}\label{commGF}
G_{-}^{\textrm{R}}(k,t)=-\frac{i}{\hbar}\:\theta(t)\langle\textrm{G}_0|
[c_k^{\dagger}(t),c_k]_{-}|\textrm{G}_0\rangle \:,
\end{equation}
is essentially equivalent in the problem at hand. Because here $|G_0\rangle = |0 \rangle$ 
is the vacuum state of the coupled e-ph system, $c_k^{\dagger}(t)c_k|\textrm{G}_0\rangle=0$, 
implying that here $G_{-}^{\textrm{R}}(k,t)=-G_{+}^{\textrm{R}}(k,t)$.

The real-space retarded (commutator) Green's functions 
\begin{equation}
G_{nn'}^{\textrm{R}}(t)\equiv -\frac{i}{\hbar}\:\theta(t)\langle
\textrm{G}_0|[c_n^{\dagger}(t),c_{n'}]_{-}|\textrm{G}_0\rangle \:,
\end{equation}
which in systems with a discrete translational symmetry depend only on $n-n'$, 
are obtained from the above momentum-space ones [cf. Eq.~\eqref{commGF}] 
via Fourier transformation:
\begin{equation}
G^{\textrm{R}}_{-}(k,t)=N^{-1}\sum_{n,n'} e^{ik\cdot (n-n')}
G_{nn'}^{\textrm{R}}(t) \:.
\end{equation}
These real-space commutator Green's functions can straightforwardly be recast 
as~\cite{Stojanovic+:14}
\begin{equation}
G_{nn'}^{\textrm{R}}(t)= -\frac{i}{\hbar}\:\theta(t)\langle\textrm{G}_0|
[\sigma_n^{+}(t),\sigma_{n'}^{-}]_{-}|\textrm{G}_0\rangle \:,
\end{equation}
where we have switched from spinless-fermion- to the pseudospin-$1/2$ operators via 
the Jordan-Wigner (JW) transformation~\cite{ColemanBOOK}:
\begin{eqnarray} \label{JWtsf}
\sigma_{n}^{z} &=& 2c_{n}^{\dagger}c_{n}-1\:, \nonumber\\
\sigma_{n}^{+} &=& 2c_{n}^{\dagger}\:e^{i\pi\sum_{l<n}c_{l}^{\dagger}c_l}\:.
\end{eqnarray}

These last Green's function can succinctly be rewritten in the form~\cite{Stojanovic+:14}
\begin{equation}\label{gnRt}
G_{nn'}^{\textrm{R}}(t)= \mathcal{G}^{xx}_{nn'}+\mathcal{G}^{yy}_{nn'}
-i(\mathcal{G}^{xy}_{nn'}-\mathcal{G}^{yx}_{nn'})\:,
\end{equation}
where $\mathcal{G}^{\alpha\beta}_{nn'}$ (\:$\alpha,\beta=x,y$\:) are retarded 
pseudospin correlation functions, defined as~\cite{Stojanovic+:14}
\begin{equation}
\mathcal{G}^{\alpha\beta}_{nn'}\equiv -\frac{i}{\hbar}\:\theta(t)\langle\textrm{G}_0|
[\sigma_n^{\alpha}(t),\sigma_{n'}^{\beta}]_{-}|\textrm{G}_0\rangle \:.
\end{equation}
[Note that, for the sake on notational simplicity, we omitted the time argument and 
the superscript $R$ in the notation for these last correlation functions.] 

\subsection{Multi-qubit Ramsey interference protocol}\label{Ramsey_subsec}
The many-body (multi-qubit) Ramsey interference protocol is applicable to systems in which addressability 
at the single-qubit (spin) level is feasible. This protocol yields the real-space and time-resolved 
commutator Green's functions of spin (or, in the case of qubits, pseudospin-$1/2$-) 
operators~\cite{Knap++:13}.

The protocol entails a special type of Rabi pulses, which can quite generally be 
parameterized as
\begin{equation}
R_n(\theta,\phi)\equiv\mathbbm{1}_{2}\cos\frac{\theta}{2} + 
i(\sigma_n^{x}\cos\phi-\sigma_n^{y}\sin\phi)\sin\frac{\theta}{2} \:,
\end{equation}
with $\theta=\Omega\tau$ (where $\Omega$ is the corresponding Rabi frequency and 
$\tau$ the duration of the pulse) and $\phi$ being the phase of the laser field. 
The special type of such pulses of relevance for the Ramsey protocol is the one 
with $\theta=\pi/2$ and an arbitrary $\phi$, which is hereafter denoted as $R_n(\phi)$.

Generally speaking, the Ramsey-interference protocol entails, as its first step, a local 
$\pi/2$-rotation at site $n$ (with the value $\phi_1$ of the parameter $\phi$); this step 
is followed by an evolution of the system over the time interval of duration $t$ and a local 
$\pi/2$-rotation at site $n'$ or global $\pi/2$-rotation (with the value $\phi_2$ of the parameter
$\phi$). As its final step, this protocol requires a measurement of $\sigma_{n'}^{z}$, i.e. 
the $z$-component of the pseudospin at site $n'$. Consequently, the final measurement result 
is given by~\cite{Knap++:13}
\begin{equation} \label{MeasureResRamsey}
M_{nn'}(\phi_1,\phi_2,t)=\langle\textrm{S}(t)
|\:\sigma_{n'}^{z}\:|\textrm{S}(t)\rangle\:,
\end{equation}
where $|\textrm{S}(t)\rangle$ is the state obtained from $|\textrm{G}_0\rangle$ 
after carrying out the first three steps of the Ramsey protocol:
\begin{equation}
|\textrm{S}(t)\rangle\equiv R_{n'}(\phi_2)U_H(t) R_n(\phi_1)
|\textrm{G}_0\rangle \:.
\end{equation}

It is straightforward to demonstrate that for a system that has the $U(1)$ symmetry 
under pseudospin rotations around the $z$ axis and the one under reflections 
with respect to the $z$ axis, the expression for the final measurement result 
in the Ramsey protocol [cf. Eq.~\eqref{MeasureResRamsey}] reduces to~\cite{Stojanovic+:14}
\begin{multline}
M_{nn'}(\phi_1,\phi_2,t)=-\frac{1}{4}\big[\sin(\phi_1-\phi_2)
(\mathcal{G}^{xx}_{nn'}+\mathcal{G}^{yy}_{nn'}) 
\\
-\cos(\phi_1-\phi_2)(\mathcal{G}^{xy}_{nn'}
-\mathcal{G}^{yx}_{nn'})\big] \:.
\end{multline}
From this last result, it can be inferred that the terms 
$\mathcal{G}^{xx}_{nn'}+\mathcal{G}^{yy}_{nn'}$ and $\mathcal{G}^{xy}_{nn'}
-\mathcal{G}^{yx}_{nn'}$ required to determine $G_{nn'}^{\textrm{R}}(t)$ [cf.
Eq.~\eqref{gnRt}] are given by $M_{nn'}(\phi_1,\phi_2,t)$ for two
different choices of the angles $\phi_1$ and $\phi_2$ (namely, for 
$\phi_1-\phi_2=\pm \pi/2$ and $\phi_1=\phi_2$, respectively).

Once the Green's functions $G_{nn'}^{\textrm{R}}(t)$ [cf. Eq.~\eqref{gnRt}] are obtained 
using the above scheme, the commutator Green's function $G_{-}^{\textrm{R}}(k,t)$ [cf.
Eq.~\eqref{commGF}] for an arbitrary quasimomentum $k$ can be determined using a spatial 
Fourier transformation, whereby its anticommutator counterpart $G_{+}^{\textrm{R}}(k,t)$ 
[cf. Eq.~\eqref{anticommGF}] is straightforward to recover. Finally, having obtained 
$G_{+}^{\textrm{R}}(k,t)$, using Eq.~\eqref{CpectFuncDef} one can compute the spectral 
function $A(k,\omega)$ for a broad range of frequencies through a numerical Fourier transform 
to the frequency domain.

\section{Simulator and its governing Hamiltonian} \label{SimulatorAndHamiltonian}
In what follows, we first briefly describe the layout of the SC analog simulator 
to be considered in the remainder of this work (Sec.~\ref{SimulatorLayout}). We follow this up 
with a detailed derivation of its underlying effective Hamiltonian and its mapping to a coupled 
e-ph model with Peierls- and breathing-mode type e-ph interactions (Sec.~\ref{EffectHamiltonian}).

\subsection{Layout of the analog simulator} \label{SimulatorLayout}
\begin{figure}[b!]
\includegraphics[clip,width=8.6cm]{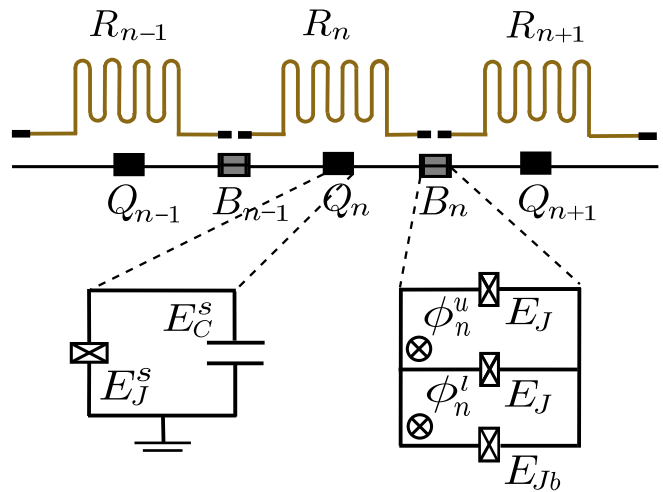}
\caption{\label{fig:circuit}(Color online)Layout of the envisioned analog simulator~\cite{Stojanovic+:14}, 
which consists of SC qubits $Q_{n}$ (with single-qubit charging- and Josephson energies $E^{s}_{C}$ 
and $E^{s}_{J}$, respectively), resonators $R_{n}$, and coupler circuits $B_{n}$. Here $\phi_{n}^{l}$ 
and $\phi_{n}^{u}$ are the respective total magnetic fluxes threading the lower and upper loops 
of $B_{n}$. }
\end{figure}

The main building blocks of the envisioned simulator, depicted in Fig.~\ref{fig:circuit}, are 
SC qubits ($Q_n$), resonators ($R_{n}$), and coupler circuits ($B_{n}$) with three JJs ($n=1,
\ldots,N$). The pseudospin-$1/2$ degree of freedom of the qubit is represented by the operators 
$\bm{\sigma}_n$, while microwave photons in the resonators, created (annihilated) by the operators 
$a_{n}^{\dagger}$ ($a_{n}$), mimic Einstein (zero-dimensional) phonons. The Hamiltonian of the 
$n$-th repeating unit of the simulator, which consists of the qubit $Q_n$ with the energy splitting 
$\varepsilon_{z}$ and the resonator $R_{n}$ with the photon frequency $\omega_{c}$, is given by
\begin{equation}\label{eq:Hn}
H_{n}^{0}=\frac{\varepsilon_{z}}{2}\:\sigma_{n}^{z}
+\hbar\omega_{c}\:a_{n}^{\dagger}a_{n}\:.
\end{equation}

Qubit $Q_{n}$ interacts with its nearest neighbors $Q_{n-1}$ and $Q_{n+1}$ through 
coupler circuits $B_{n-1}$ and $B_n$, which represent a generalization of a SQUID loop. 
The total energy of $B_{n}$ is given by 
\begin{equation}\label{JosephsonEn}
H_{n}^{J}=-\sum_{i=1}^{3}E^{i}_{J}\cos\varphi_{n}^{i} \:,
\end{equation}
where $\varphi_{n}^{i}$ ($i=1,2,3$) are the phase drops on the three JJs and $E^{i}_{J}$ their 
respective Josephson energies; it will hereafter be assumed that junctions $1$ 
and $2$ have the same energy (i.e. $E^{1}_{J}=E^{2}_{J}\equiv E_{J}$), while the third one is
different from the first two (i.e. $E^{3}_{J}=E_{Jb}\neq E_{J}$). The qubit- and resonator degrees 
of freedom are mutually coupled in this system through the flux of the resonator modes piercing the 
upper loops of coupler circuits~\cite{Lagemann+:22}. Apart from this inductive-coupling mechanism, 
those circuits are also driven by a microwave radiation (ac flux) and with an external dc flux
through their lower loops. 

In the following, $\phi_{n}^{u}$ and $\phi_{n}^{l}$ will be used to denote the 
total magnetic fluxes in the upper- and lower loops of $B_{n}$, respectively. [Note 
that hereafter all the fluxes in the problem will be expressed in units 
of $\Phi_{0}/2\pi$, with $\Phi_{0}\equiv hc/(2e)$ being the flux quantum.] 
The flux $\phi_{n}^{u}$ in the upper loop includes ac-driving contribution $\pi\cos(\omega_{0}t)$
and a dynamically-fluctuating one from the resonator modes $a_n$ and $a_{n+1}$; the latter is 
proportional to the difference of the photon displacement fields of resonators $n+1$ and $n$,
i.e. $\phi_{n,\textrm{res}}=\delta\theta_r[(a_{n+1}+a_{n+1}^{\dagger})-(a_{n}+a_{n}^{\dagger})]$, 
with the constant of proportionality $\delta\theta_r$ whose value depends on the geometric properties
of the specific resonator~\cite{Stojanovic+:14}. The lower-loop 
counterpart $\phi_{n}^{l}$ of $\phi_{n}^{u}$ also includes an ac contribution, which is given 
by $-(\pi/2)\cos(\omega_{0}t)$. Finally, the lower-loop flux is also assumed to have a dc part 
$\phi_{\textrm{dc}}$. This dc flux constitutes -- aside from $\omega_{0}$ -- the only tunable 
parameter in this system, hence being the principal experimental knob. 

\subsection{Effective system Hamiltonian} \label{EffectHamiltonian}
Due to the explicit time dependence of the ac-driving, it is favorable to switch to the rotating 
frame of the drive. This change of frames not only leads to a shift in the resonator 
(photon) frequency ($\omega_c\rightarrow \delta\omega\equiv\omega_c-\omega_0$) but also causes the 
Josephson-coupling term to acquire a time dependence. However, it can straightforwardly be demonstrated 
that this time dependence can safely be disregarded due to its rapidly-oscillating character, i.e. 
based on the rotating-wave approximation (RWA). After dropping those terms, the remaining part of 
the Josephson-coupling term is given by
\begin{equation}\label{eq:HJn_final}
\bar{H}_{n}^{J}=-E_n^{J}\cos(\varphi_{n}-\varphi_{n+1})  \:,
\end{equation}
where $\varphi_{n}$ is the gauge-invariant phase variable corresponding to the SC island of the $n$-th 
qubit~\cite{VoolDevoretReview:17} and 
\begin{equation}\label{eq:EnJ}
E_n^{J}=E_{Jb}(1+\cos\phi_{\textrm{dc}})- 
E_{J}J_1(\pi/2)\phi_{n,\textrm{res}}\:,
\end{equation}
where $J_n(x)$ are Bessel functions of the first kind. The specific form of the last equation result 
from the assumption that $E_{Jb}$ is chosen, without significant loss of generality, to be given by
$2E_{J}J_{0}(\pi/2)$. 

In the relevant regime for transmon qubits ($E^{s}_{J}\gg E^{s}_{C}$, where $E^{s}_{C}$ and $E^{s}_{J}$ 
are the charging- and Josephson energies of a single qubit, respectively) it is permissible to expand 
$\cos(\varphi_{n}-\varphi_{n+1})$ up to the second order in $\varphi_{n}-\varphi_{n+1}$, where this expansion 
is controlled by the small parameter $\delta\varphi_{0}^{2}\equiv (2E^{s}_{C}/E^{s}_{J})^{1/2}$ 
(the quantum displacement of the gauge-invariant phase); for a typical transmon qubit ($E^{s}_{J}/E^{s}_{C}\sim 100$)
one finds $\delta\varphi_{0}^{2}\sim 0.15$. Importantly, higher powers of $\varphi_{n}-\varphi_{n+1}$ 
in that expansion can be neglected not only because of the smallness of this phase difference, but also 
due to the rapidly decreasing coefficients in the expansion, which are proportional to higher powers 
of $\delta\varphi_{0}^{2}$. By switching to the pseudospin-$1/2$ operators $\bm{\sigma}_n$, $\cos(\varphi_{n}
-\varphi_{n+1})$ can be rewritten (up to an additive constant, which is irrelevant for our present purposes) 
in the form [for a detailed derivation, see Appendix~\ref{DerivCosform}]
\begin{eqnarray} \label{cosform}
\cos(\varphi_{n}-\varphi_{n+1}) &\approx& \delta\varphi_{0}^{2} 
\Big[\sigma_{n}^{+}\sigma_{n+1}^{-}+\sigma_{n}^{-}\sigma_{n+1}^{+} \nonumber\\
&-& \frac{\sigma_{n}^{z}+\sigma_{n+1}^{z}}{2}\Big] \:,
\end{eqnarray}
The first term on the RHS of the last equation corresponds to an $XY$-type coupling between two adjacent transmons, 
while the second one -- being of the same form as the $\sigma_{n}^{z}$ term in Eq.~\eqref{eq:Hn} -- describes a shift 
in their (single-qubit) frequency.

It is worthwhile mentioning that in the derivation of Eq.~\eqref{cosform} the terms 
of the type $\sigma_{n}^{+}\sigma_{n+1}^{+}$ and $\sigma_{n}^{-}\sigma_{n+1}^{-}$ have been neglected. 
While this is conventionally done -- by virtue of the RWA -- even in the most general (multilevel) treatment 
of transmons, in the problem at hand there is an even more rigorous argument for doing so. Namely,
here we are concerned with a single-excitation polaron problem. Therefore, the part of the total
Hilbert space of relevance for this problem consists of states with a single spinless fermion, which 
in the pseudospin-$1/2$ (qubit) language translates into states with precisely one qubit in the 
logical state $|1\rangle$. Accordingly, the terms $\sigma_n^- \sigma_{n+1}^-$ and $\sigma_n^+ 
\sigma_{n+1}^+$ both yield zero when acting on an arbitrary state in the relevant part of the Hilbert 
space of the system.

At this point it is pertinent to recast the problem at hand in terms of the spinless-fermion
operators $\{c_{n}, c_{n}^{\dagger}\}$ (instead of the pseudospin-$1/2$ operators $\bm{\sigma}_n$) 
via the JW transformation [cf. Eq.~\eqref{JWtsf}], whereby $\sigma_{n}^{+}\sigma_{n+1}^{-}+\sigma_{n}^{-}
\sigma_{n+1}^{+} \rightarrow c_{n}^{\dagger}c_{n+1}+\text{H.c.}$. This allows us to write the 
effective system Hamiltonian in the form of a lattice model that describes an itinerant spinless-fermion 
excitation interacting with zero-dimensional bosons (phonons) through two different e-ph coupling 
mechanisms.

The noninteracting (free) part $H_{0}$ of the effective system Hamiltonian consists 
of the free-excitation (hopping) and free-phonon terms
\begin{equation}\label{nonint}
H_{0}=-t_{0}(\phi_{\textrm{dc}})\sum_{n}(c_{n}^{\dagger}c_{n+1}
+\textrm{H.c.})+\hbar\delta\omega\sum_{n}a_{n}^{\dagger}a_{n}\:,
\end{equation}
where $t_{0}(\phi_{\textrm{dc}})\equiv E_{Jb}\delta\varphi_{0}^{2}(1+\cos\phi_{\textrm{dc}})$ plays the
role of the effective bare-excitation hopping amplitude. This hopping amplitude can be tuned 
by varying the dc flux $\phi_{\textrm{dc}}$. [Strictly speaking, $H_{0}$ also contains
diagonal terms $c_{n}^{\dagger}c_{n}$, which originate from the terms with $\sigma_{n}^{z}$
in Eq.~\eqref{eq:Hn}, as well as from the expansion of $\cos(\varphi_{n}-\varphi_{n+1})$. Yet,
these terms can be disregarded as they only lead to a band offset for spinless fermions.]
On the other hand, the interacting part includes two e-ph coupling terms and is given by~\cite{StojanovicPRL:20}
\begin{eqnarray}\label{Heph} 
H_{\mathrm{e-ph}} &=& g\hbar\delta\omega\:l_0^{-1}\sum_{n}\Big[(c_{n}^{\dagger}
c_{n+1}+\textrm{H.c.})\:\left(u_{n+1}-u_{n}\right) \nonumber \\
&-& c_{n}^{\dagger}c_{n}\left(u_{n+1}-u_{n-1}\right)\Big]\:,  
\end{eqnarray}%
where $g$ is the dimensionless e-ph coupling strength and $u_n \equiv l_0(a_{n}+a_{n}^{\dagger})$, 
with $l_0$ being the zero-point length of the Einstein oscillator with frequency $\delta\omega$. 
The first term corresponds to the Peierls-coupling mechanism, which accounts for the lowest-order
(linear) dependence of the effective (phonon-modulated) hopping amplitude between sites $n$ and $n+1$ 
on the difference $u_{n+1}-u_n$ of the corresponding phonon displacements~\cite{Stojanovic+:04}. 
The second term corresponds to the breathing-mode type coupling~\cite{Slezak++:06}, i.e. the 
antisymmetric coupling of the excitation density $c_{n}^{\dagger}c_{n}$ at site $n$ with the 
phonon displacements on sites $n\pm 1$.

When recast in momentum space, $H_{\mathrm{e-ph}}$ assumes the form 
$N^{-1/2}\sum_{k,q}\gamma_{\textrm{e-ph}}(k,q)\:c_{k+q}^{\dagger}c_{k}
(a_{-q}^{\dagger}+a_{q})$, where
\begin{equation}\label{PBMvertexFunc}
\gamma_{\textrm{e-ph}}(k,q)=2ig\hbar\delta\omega\:[\:\sin k+\sin q-\sin(k+q)]
\end{equation}
is the corresponding e-ph vertex function (note that here quasimomenta are expressed in units of the 
inverse lattice constant, thus the quasimomenta from the Brillouin zone belong to $(-\pi,\pi]$). For 
the most general (momentum-dependent) vertex function $\gamma_{\textrm{e-ph}}(k,q)$, the effective
e-ph coupling strength is given by 
\begin{equation} \label{genlambda}
\lambda_{\textrm{eff}}=\frac{\langle|\gamma_{\textrm{e-ph}}(k,q)|^{2}\rangle}
{2t_{\rm e}\:\hbar\omega_{\textrm{ph}}}\:, 
\end{equation}
where $t_{\rm e}$ is the excitation hopping amplitude, $\omega_{\textrm{ph}}$ the phonon frequency, 
and $\langle\ldots\rangle$ stands for the Brillouin-zone average over quasimomenta $k$ and $q$:
\begin{equation}\label{}
\langle|\gamma_{\textrm{e-ph}}(k,q)|^{2}\rangle \equiv\frac{1}{(2\pi)^2}
\:\int^{\pi}_{-\pi}\int^{\pi}_{-\pi}\:|\gamma_{\textrm{e-ph}}(k,q)|^{2}\:dkdq  \:.
\end{equation}
In the problem at hand -- where $t_{\rm e}\rightarrow\:t_0$, $\omega_{\textrm{ph}}\rightarrow
\:\delta\omega$, and the vertex function is given by Eq.~\eqref{PBMvertexFunc} -- the effective 
coupling strength evaluates to $\lambda_{\textrm{eff}}\equiv 3g^{2}\:\hbar\delta\omega/t_{0}$
and depends on $\phi_{\textrm{dc}}$.

The fact that the vertex function in Eq.~\eqref{PBMvertexFunc} depends both on $q$ and $k$ implies 
that the Hamiltonian $H_{\textrm{eff}}=H_{0}+H_{\textrm{e-ph}}$ does not belong to the realm 
of applicability of the Gerlach-L\"{o}wen theorem~\cite{Gerlach+Lowen:87}, which rules out a nonanalytic 
behavior of ground-state-related quantities. It was already demonstrated that the ground state of this 
Hamiltonian displays a sharp (level-crossing) transition at a critical value of the effective coupling 
strength (tuned here by varying $\phi_{\textrm{dc}}$)~\cite{Stojanovic+:14}. Whereas below the critical 
value the system has zero-quasimomentum ground state -- that is, the energy minimum corresponds to the 
eigenvalue $K=0$ of the total quasimomentum operator 
\begin{equation}
K_{\mathrm{tot}}=\sum_{k}k\:c_{k}^{\dagger}c_{k}+\sum_{q}q\:a_{q}^{\dagger}a_{q} 
\end{equation}
above this critical value $H_{\textrm{eff}}$ has a twofold-degenerate ground state. This unconventional, 
degenerate ground state corresponds to the pair of equal and opposite (nonzero) quasimomenta $\pm K_{\textrm{gs}}$, 
where $K_{\textrm{gs}}$ reaches the value of $\pi/2$ for a sufficiently strong coupling (i.e. sufficiently 
large $\lambda_{\textrm{eff}}$).

It is worthwhile pointing out that the system at hand has the peculiar property that the $k=0$ Bloch state 
$|\Psi_{k=0}\rangle \equiv c^{\dagger}_{k=0}|0\rangle_{\textrm{e}}\otimes|0\rangle_{\textrm{ph}}$ of a bare 
excitation is an exact eigenstate of $H_{\textrm{eff}}$ for an arbitrary coupling strength, regardless of the 
value of $\lambda_{\textrm{eff}}$. This fact has profound consequences for the resulting spectral function (cf. 
Sec.~\ref{resdiscuss} below). Moreover, for effective coupling strengths $\lambda_{\textrm{eff}}$ below the 
critical one $|\Psi_{k=0}\rangle$ is the lowest-energy eigenstate of $H_{\textrm{eff}}$ , i.e., its ground state.
The state $|\Psi_{k=0}\rangle$, when recast in terms of pseudospin-$1/2$ (qubit) degrees of freedom via the
JW transformation [cf. Eq.~\eqref{JWtsf}], corresponds to an $N$-qubit $W$ state~\cite{StojanovicPRL:20}; 
more generaly, the bare-excitation Bloch state $|\Psi_{k}\rangle$ with quasimomentum $k$ corresponds to a 
twisted $N$-qubit $W$ state~\cite{Haase++:22}.

The dimensionless e-ph coupling strength $g$ [cf. Eq.~\eqref{Heph}] is given by $g\hbar\delta\omega=\delta
\varphi_{0}^{2}\:E_{J}J_1(\pi/2)\delta\theta_r$. While $g$ itself does not depend on $\omega_0$ and 
$\phi_{\textrm{dc}}$, $\lambda_{\textrm{eff}}$ inherits the dependence on the latter parameter from $t_{0}$ 
and can thus be tuned by varying $\phi_{\textrm{dc}}$. More precisely, the effective e-ph coupling strength 
has the following dependence on the dc flux:
\begin{equation}\label{expr_lambda}
\lambda_{\textrm{eff}}(\phi_{\textrm{dc}})=\frac{3}{2}\:g\:\frac{J_1(\pi/2)\delta\theta_r}
{J_0(\pi/2)\left(1+\cos\phi_{\textrm{dc}}\right)} \:.
\end{equation}
For a typical SC microwave resonator the value of $\delta\theta_r$ is around $3.5\times 10^{-3}$, while the effective 
phonon frequency $\delta\omega$ can be taken to be in the range $\delta\omega/2\pi=200 - 300$ MHz. By choosing the 
Josephson energy $E_{J}$ such that the condition $\delta\varphi_{0}^{2}\:E_{J}/2\pi\hbar=100$ GHz is satisfied, one 
obtains $g\delta\omega/2\pi=198$\:MHz. 

The system under consideration allows one to access both the adiabatic- ($t_{0}>\hbar\delta\omega$) 
and antiadiabatic regime ($t_{0}<\hbar\delta\omega$) of small-polaron physics by varying $\phi_{\textrm{dc}}/\pi$ 
in the fairly narrow range $0.95-0.99$. For the choice $\delta\omega/2\pi=200$\:MHz of the effective phonon 
frequency, for instance, the onset of the antiadiabatic regime is for $\phi_{\textrm{dc}}/\pi\approx 0.980$. 
For the same choice of parameters, the sharp ground-state transition takes place for $\phi_{\textrm{dc}}/\pi\approx 0.968$,
the corresponding value of the effective coupling strength $\lambda_{\textrm{eff}}$ being approximately equal 
to $1.24$. Importantly, in the same narrow range of values for $\phi_{\textrm{dc}}$, $\lambda_{\textrm{eff}}$ can assume 
both values within the weak-coupling regime (e.g., for $\delta\omega/2\pi=200$ MHz and $\phi_{\textrm{dc}}
/\pi=0.95$, one finds the effective coupling strength $\lambda_{\textrm{eff}}=0.51$) and in the strong-coupling one 
(e.g., for $\delta\omega/2\pi=200$ MHz and $\phi_{\textrm{dc}}/\pi=0.98$ one obtains $\lambda_{\textrm{eff}}=3.17$).

Both bare-excitation- and dressed-excitation Bloch states can be prepared in the system at hand, starting 
from the initial state $|G_0\rangle$, using a microwave-driving protocol proposed in Ref.~\cite{Stojanovic+:14}.
This protocol, which is based on the discrete translational symmetry of the system and energy conservation, 
allows the preparation of the desired Bloch states within times $3-4$ orders of magnitude shorter than the 
currently achievable decoherence times $T_2$ of SC qubits.

\section{Results and Discussion}     \label{resdiscuss}
Using the KPM (for the essential aspects of this computational scheme, see Appendix~\ref{KPMforSpectFunc}) the single-particle 
spectral function [cf. Eq.~\eqref{MomFreqSpectFunc}] was evaluated for a system with $N=10$ sites and the total of 
$N^{\textrm{max}}_{\textrm{ph}}=18$ phonons in the truncated phonon Hilbert space; the dimension of the latter Hilbert 
space is $D_{\textrm{ph}}= 43,758$ [for general aspects of the Hilbert-space truncation, see Appendix~\ref{ExactDiag}]
In order to achieve a good resolution, we evaluated as many as $10^5$ Chebyshev moments [cf. Appendix~\ref{KPMforSpectFunc}] 
in the expansion of the desired spectral function $A(k,\omega)$.

Our KPM-based evaluation of the momentum-frequency resolved spectral function was carried out on a $64$-core, 
$7$\:GHz AMD Ryzen Threadripper PRO 5995WX workstation, with a total of $527$\:GB of main memory. The runs 
that were needed to obtain the results presented in this section consumed around $50$ hours. 

\begin{figure}[t!]
\includegraphics[clip,width=8.75cm]{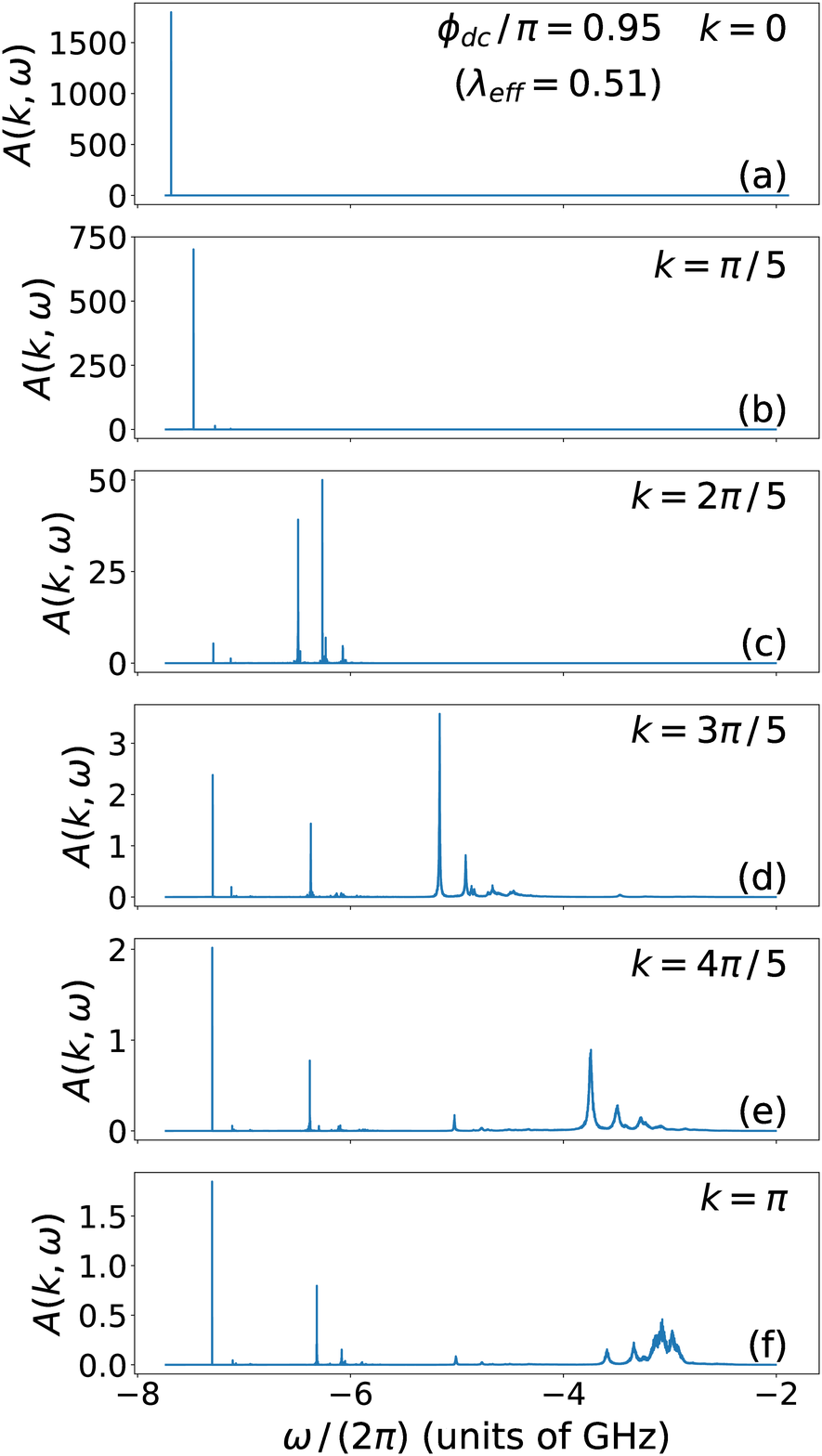}
\caption{\label{fig:SF1} Momentum-frequency resolved spectral function $A(k,\omega)$, for six different quasimomenta $k$
in the Brillouin zone, evaluated for a range of frequencies. The parameter values used are $\delta\omega/(2\pi) = 200$\:MHz
and $\phi_{\textrm{dc}}/\pi = 0.95$; the corresponding effective coupling strength is $\lambda_{\textrm{eff}}=0.51$. 
The obtained spectral function satisfies the sum rule in Eq.~\eqref{SumRuleSpectFunc}.}
\end{figure}

\begin{figure}[t!]
\includegraphics[clip,width=8.75cm]{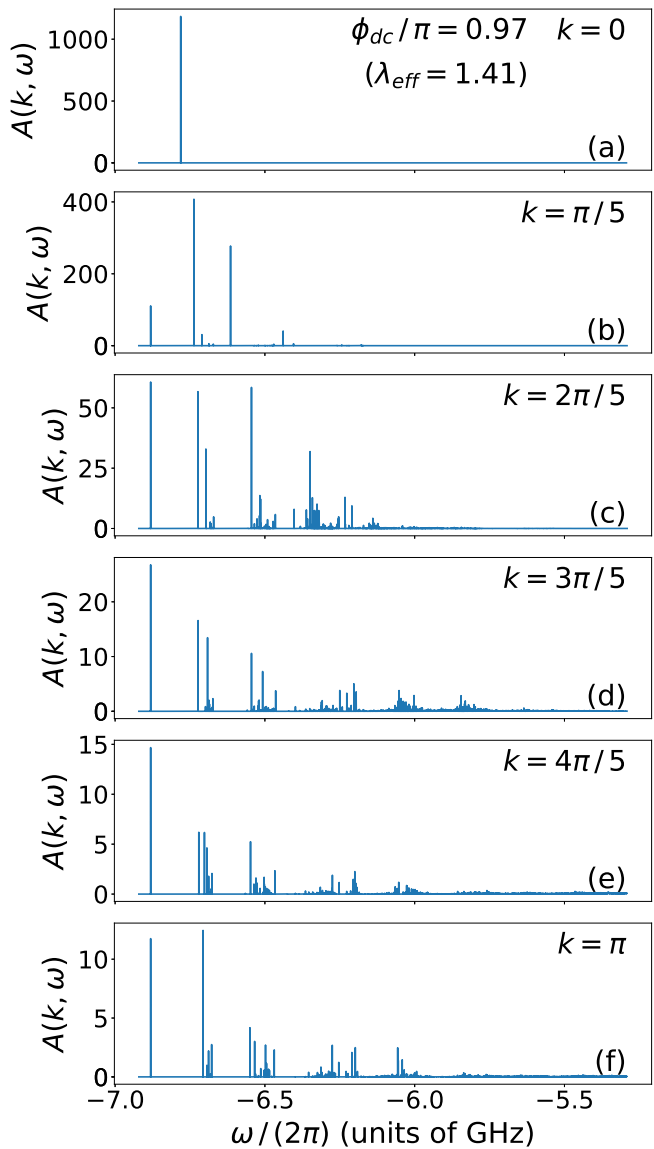}
\caption{\label{fig:SF2} Momentum-frequency resolved spectral function $A(k,\omega)$, for six different quasimomenta $k$
in the Brillouin zone, evaluated for a range of frequencies. The parameter values used are $\delta\omega / (2\pi) = 200$\:MHz
and $\phi_{\textrm{dc}}/\pi = 0.97$; the corresponding effective coupling strength is $\lambda_{\textrm{eff}}=1.41$. 
The obtained spectral function satisfies the sum rule in Eq.~\eqref{SumRuleSpectFunc}.}
\end{figure}

\begin{figure}[t!]
\includegraphics[clip,width=8.75cm]{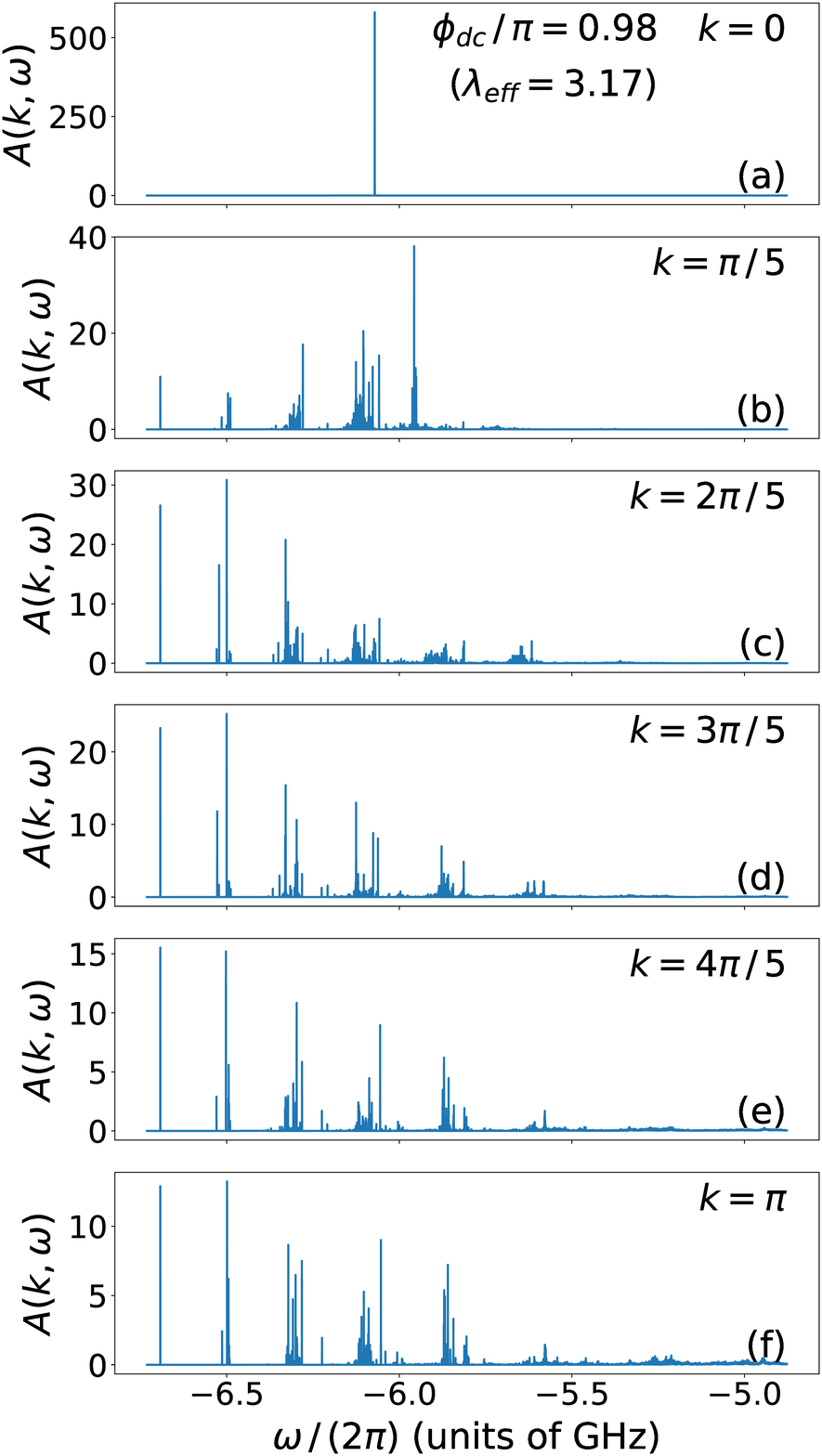}
\caption{\label{fig:SF3} Momentum-frequency resolved spectral function $A(k,\omega)$, for six different quasimomenta $k$
in the Brillouin zone, evaluated for a range of frequencies. The parameter values used are $\delta\omega/(2\pi)=200$\:MHz
and $\phi_{\textrm{dc}}/\pi = 0.98$; the corresponding effective coupling strength is $\lambda_{\textrm{eff}}=3.17$. 
The obtained spectral function satisfies the sum rule in Eq.~\eqref{SumRuleSpectFunc}.}
\end{figure}

The evaluation was carried out for three different values of the dc flux $\phi_{\textrm{dc}}$, the main experimental
knob in the system, two of which belong to the antiadiabatic regime ($\phi_{\textrm{dc}}/\pi = 0.95$ and $0.97$) and 
one to the adiabatic one ($\phi_{\textrm{dc}}/\pi = 0.98$). The resulting frequency dependence of the 
spectral function for six different quasimomenta ($k=0,\:\pi/5,\:2\pi/5,\:3\pi/5,\:4\pi/5,\:\pi$) in the positive half 
of the Brillouin zone, consistent with periodic boundary conditions, is depicted in Figs.~\ref{fig:SF1} - \ref{fig:SF3}.
In each particular case, it was numerically verified that the obtained spectral function satisfies the sum rule 
in Eq.~\eqref{SumRuleSpectFunc}.

In order to understand the obtained results, it is useful to recall some general properties of the energy
spectra of models describing a short-range coupling of an itinerant excitation with dispersionless
phonons; the strong-coupling regime of such models is characterized by the presence of heavily-dressed excitations
(small polarons). Regardless of the specific form of the e-ph interaction, the center of the small-polaron 
Bloch band is situated at an energy $E_b$ below that of a bare excitation, this last energy being referred 
to as the small-polaron binding energy.

At a fixed quasimentum $k$, the sum over eigenstates that contributes to the spectral function includes
the discrete states (i.e. those that belong to coherent polaron Bloch bands) and their respective continua;
importantly, the energetic width of each of those continua is equal that of the respective polaron Bloch band.
In particular, the one-phonon continuum represents the inelastic-scattering threshold -- i.e. the minimal 
energy that a phonon-dressed excitation ought to have to be capable of emitting a single phonon. This one-phonon 
continuum sets in at the energy $\hbar\omega_{\textrm{ph}}$ above the ground-state energy, where $\omega_{\textrm{ph}}$ 
is the phonon frequency~\cite{Engelsberg+Schrieffer:63} (recall that in the system at hand the role of the 
effective phonon freqency is played by $\delta\omega$). For a sufficiently weak e-ph coupling, a coupled e-ph 
system only has one discrete  Bloch state $|\psi^{(j=0)}_k\rangle$ at quasimomentum $k$ and its corresponding 
continuum of states pertains to a dressed excitation with quasimomentum $k-q$ and an unbound phonon with 
quasimomentum $q$. 

As the e-ph coupling strength is increasing, additional coherent polaron bands -- i.e., additional discrete states
at each quasimomentum $k$ in the Brillouin zone -- start to emerge below the aforementioned one-phonon continuum. 
The first such excited polaron state at quasimomentum $k$ -- split off from the continuum -- corresponds to a polaron 
bound with an additional phonon, their total quasimomentum being equal to $k$. For even stronger e-ph coupling there 
is another, second excited state, which represents a bound state of a polaron and two additional phonons (again, 
with the same total quasimomentum $k$). Those states, along with their respective continua, provide additional 
contributions to the spectral function $A(k,\omega)$~\cite{Fehske+:06}.

For the lowest value used ($0.95$) of the parameter $\phi_{\textrm{dc}}/\pi$ the system is in the weak-coupling
regime -- below the sharp ground-state transition. As pointed out in Sec.~\ref{EffectHamiltonian} its ground state 
corresponds to a bare-excitation Bloch state at quasimomentum $k=0$, which explains the presence of a single discrete 
peak of the spectral function $A(k=0,\omega)$ in Fig.~\ref{fig:SF1}. For larger quasimomenta $k$ new peaks, which 
correspond to excited dressed-excitation (polaron) Bloch states, gradually apear. 

In keeping with general characteristics of energy spectra of coupled e-ph systems, for increasing values of 
$\phi_{\textrm{dc}}/\pi$, which in the system under consideration translate into a larger value of the effective 
e-ph coupling strength [cf. Eq.~\ref{expr_lambda}], one can notice an increasing number of discrete peaks. In 
particular, it can be inferred from Figs.~\ref{fig:SF2} and \ref{fig:SF3} that up to the largest effective coupling 
strength considered ($\lambda_{\textrm{eff}}=3.17$) there are up to five such peaks, accompanied by their 
corresponding continua.

The salient feature of the obtained results for the spectral function at $k=0$ for all four values of 
$\phi_{\textrm{dc}}/\pi$ is the absence of the one-phonon continuum. This, seemingly peculiar, result is 
a direct consequence of the fact that -- due to equal Peierls- and breathing-mode coupling strengths in the
system at hand -- the free-excitation Bloch state with quasimomentum $k=0$ represents an exact eigenstate 
of the total system Hamiltonian at an arbitrary coupling strength [cf. Sec.~\ref{EffectHamiltonian}]. In 
other words, while only for $\phi_{\textrm{dc}}/\pi$ below the critical value of around $0.968$ this state 
is the ground state of the system, it represents an eigenstate even above the transition point where the 
ground state itself no longer corresponds to the total quasimomentum $K=0$ but instead to a pair of nonzero 
quasimomenta $\pm \pi/2$. This last circumstance also explains why in Figs.~\ref{fig:SF2} and ~\ref{fig:SF3} 
the only discrete peak for $k=0$ [part (a) of the respective figure] appears at a higher energy than the 
lowest-lying peaks at other quasimomenta [parts (b)-(f) of the respective figure].

Having described the obtained results for the spectral function, it is useful to point out the connection between 
the computed spectral properties and nonequilibrium dynamics of the coupled e-ph system~\cite{Dorfner++:15}. 
As mentioned above [cf. Sec.~\ref{genspectfunc}], the spectral function $A(k,\omega)$ is simply related -- by 
a Fourier transform in time -- to the Loschmidt amplitude $\langle\psi(t)|c^{\dagger}_k |0\rangle$ for the system 
to remain at time $t$ in the initial, bare-excitation state with quasimomentum $k$. In a typical situation -- with 
a few discrete delta-like peaks (i.e. Bloch bands of a dressed excitation) accompanied by their respective continua, 
approximately represented as Lorentzians -- this can allow one to get a qualitative picture of the system dynamics. 

Namely, knowing that the Fourier transform to the time domain of a frequency-space Lorentzian is the bilateral 
exponential function $e^{-|t|/\tau}$ [where $\tau$ is the inverse half width at half maximum (HWHM) of the 
Lorentzian], one can straightforwardly find the survival probability of the original bare-excitation state 
with quasimomentum $k$ (the Loschmidt echo) $\mathcal{L}_k(t)$ at time $t$ [cf. Eq.~\eqref{LoschmidtEcho}]. 
For instance, in the simplest case of a single sharp peak and its corresponding continuum, the latter is given 
by a sum of a constant (time-independent) term, an exponentially-decaying one, and an exponentially-damped 
oscillatory term. In cases with multiple peaks and their respectivee continua, the resulting expression for 
the Loschmidt echo is more complicated, as it involves multiple oscillatory terms, but is still straightforward 
to derive. 

\begin{figure}[t!]
\includegraphics[clip,width=8.75cm]{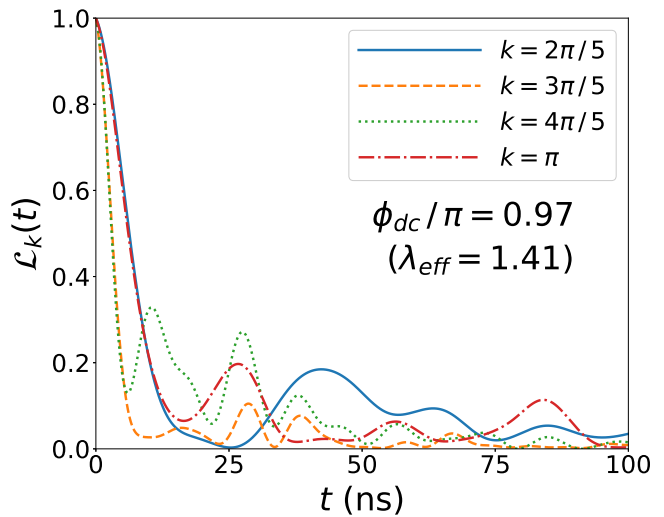}
\caption{\label{fig:LE}Time dependence of the Loschmidt echo for four different initial free-excitation 
quasimomenta ($k=2\pi/5,\:3\pi/5,\:4\pi/5$, and $\pi$). The parameter values used are the same as in 
Fig.~\ref{fig:SF2}.}
\end{figure}

The typical decaying oscillatory behavior of the Loschmidt echo is illustrated in Fig.~\ref{fig:LE}. 
The rapid decay of this quantity as a function of time $t$ after an e-ph interaction quench reflects the fact 
that the initially bare itinerant excitation becomes increasingly phonon-dressed over time. On the other hand, 
the oscillatory features of this quantity originates from the fact that the initial bare-excitation Bloch 
state $c^{\dagger}_k|0\rangle$ at a generic quasimomentum $k$ is not an eigenstate of the coupled e-ph Hamiltonian 
of the system (i.e. the Hamiltonian describing the post-quench dynamics), but is rather given by a linear 
combination of multiple eigenstates of this Hamiltonian (recall, however, that the system at hand has a rather 
unconventional property that the $k=0$ bare-excitation Bloch state is its exact eigenstate for an arbitrary e-ph 
coupling strength; consequently, the dynamics for $k=0$ are completely trivial). This oscillatory behavior of the 
Loschmidt echo can be recovered more accurately by directly solving the Schr\"{o}dinger equation for a phonon-dressed 
excitation using a Chebyshev expansion of its time-evolution operator~\cite{Stojanovic+Salom:19}.

For completeness, it is of interest to comment on the capability of the proposed SC
system for accurate measurements of the momentum-frequency resolved spectral function of dressed excitations in the underlying 
coupled e-ph model (cf. Sec.~\ref{genspectfunc}) and, by extension, elucidating quantum dynamics of the small-polaron formation.
The actual resolution with which the retarded single-particle Green's function can be determined using the many-body version 
of the Ramsey interference protocol (cf. Sec.~\ref{Ramsey_subsec}) is certainly setup-specific, i.e. dependent on the specific 
data-acquisition capabilites available. Accordingly, the frequency resolution that can be achieved in an experimental realization 
of our proposal for measuring the spectral function will also be dependent on the specific setup. However, a general 
argument can be provided as to why accurate measurements of the spectral function -- as well as drawing quantitative 
conclusions about the small-polaron formation dynamics -- in the proposed system is conceivable with existing technology.

Namely, the characteristic energy scales in the proposed SC analog simulators are 
$3 - 4$ orders of magnitude smaller than their counterparts in solid-state systems. For instance, 
the frequencies of microwave photons that emulate phonons in this system are of the order of $0.1$\:GHz, 
while a typical optical-phonon frequency in the solid state is of the order of a few THz. As a result, 
the dynamics of our proposed system are $3 - 4$ orders of magnitude slower than that of its solid-state 
counterparts, which -- in fact -- is the principal reason as to why such analog simulators can be put 
to good use in elucidating complex dynamical processes. For this reason, assuming similar technological
capabilities in both cases, we can conclude that it should actually be more straightforward to measure 
the relevant dynamical response functions accurately in the proposed synthetic SC system than in 
naturally-occurring electronic materials.

\section{Summary and Conclusions}     \label{sumconcl}
In this paper, we proposed a scheme for investigating spectral properties of polaronic excitations using a 
superconducting analog simulator based on an array of inductively-coupled superconducting qubits and 
microwave resonators. This system emulates a model describing an itinerant spinless-fermion excitation coupled 
to dispersionless phonons via Peierls- and breathing-mode type excitation-phonon interactions. Using the kernel 
polynomial method we computed the momentum-frequency resolved spectral function of this system for several different 
choices of its parameters, covering both the adiabatic and antiadiabatic regimes of this model. In addition, 
we indicated how this spectral function can be recovered experimentally using the many-body version of the Ramsey 
interference protocol.

We found implications of strong, nonlocal excitation-phonon coupling in the investigated system for its 
spectral properties. A salient feature of this system -- resulting from the fact that Peierls' and breathing-mode
coupling strengths in this system are equal -- is that its single-particle spectral function does not show
a one-phonon continuum at zero quasimomentum. This is a direct implication of the fact that the bare-excitation
Bloch state with zero-quasimentum is an exact eigenstate of the coupled excitation-phonon Hamiltonian of the
system for an arbitrary excitation-phonon coupling strength.

The present study of spectral properties can be extended to other types of analog simulators~\cite{HuangEtAlNJP:21}. 
Namely, essentially the same coupled e-ph model can be experimentally realized as an analog simulator 
based on an array of neutral atoms in optical tweezers that interact through Rydberg-dressed resonant 
dipole-dipole ineraction~\cite{StojanovicPRA:21}. Therefore, the same experimental investigation of 
spectral properties of polarons born 
out of the interplay of Peierls' and breathing-mode type e-ph interactions can also be carried
out using that atomic system, with the added benefit that this last system allows one to independently
vary both coupling strengths

\begin{acknowledgments}
This research was supported by the Deutsche Forschungsgemeinschaft (DFG) -- SFB 1119 -- 236615297.
\end{acknowledgments}

\appendix 

\section{Derivation of Eq.~(\ref{cosform})}  \label{DerivCosform}
In what follows, we provide the derivation of the approximate expression for $\cos(\varphi_n -\varphi_{n+1})$
[Eq.~\eqref{cosform} in Sec.~\ref{EffectHamiltonian}] in the regime of interest for transmon qubits.

We start from the quantization of the gauge-invariant phase variables $\varphi_n$ in terms of bosonic 
operators~\cite{Girvin:14}
\begin{equation}\label{QuantPhase}
\varphi_n = \delta\varphi_{0} (b_n + b_n^{\dagger}) \:,
\end{equation}
where $\delta\varphi_{0} \equiv (2E^{s}_{C}/E^{s}_{J})^{1/4}$. In the regime of relevance for transmons
($E^{s}_{J}/E^{s}_{C}\sim 100$), it is pertinent to expand $\cos(\varphi_n - \varphi_{n+1})$ to the second 
order in the phase difference $\varphi_n - \varphi_{n+1}$, where $\delta\varphi_{0}$ plays the role the small 
parameter controlling the expansion. In this manner, by also making use of Eq.~\eqref{QuantPhase}, we readily 
obtain
\begin{eqnarray}
\cos(\varphi_n - \varphi_{n+1}) &\approx& 1-\frac{1}{2}\:\delta\varphi^{2}_{0} 
\big[(b_n + b_n^{\dagger})^2 \nonumber \\
&-& (b_{n+1} + b_{n+1}^{\dagger})^2\big] \:.
\end{eqnarray}
By exploiting bosonic commutation relations for the operators $b_n$, $b_n^{\dagger}$, 
$b_{n+1}$, and $b_{n+1}^{\dagger}$, we further find
\begin{eqnarray} \label{ExprCosine1}
\cos(\varphi_n - \varphi_{n+1}) &\approx& 1-\frac{\delta\varphi^{2}_{0}}{2}\:
\big( b_n^{\dagger}b_n + b_{n+1}^{\dagger}b_{n+1} + 1\big)  \nonumber \\
&+& \frac{\delta\varphi^{2}_{0}}{2}\:\big(b_n b_{n+1} + b_n b_{n+1}^{\dagger} \nonumber \\
&+& b_{n}^{\dagger}b_{n+1} + b_{n}^{\dagger} b_{n+1}^{\dagger}\big) \:.
\end{eqnarray}

While up to this point the transformations were completely general, i.e. consistent with the multilevel 
treatment of transmon qubits, we now resort to the two-level approximation and switch from the bosonic 
operators $b_{n}$, $b_{n}^{\dagger}$ to the pseudospin-$1/2$ operators $\bm{\sigma}_{n}$ representing 
transmons; in the two-level approximation the bosonic occupation numbers are constrained to $0$ and $1$.
This transformation from constrained bosonic- to pseudospin operators can be seen as an inverted version 
of the Holstein-Primakoff transformation in solid-state physics~\cite{Holstein+Primakoff:40}. In the 
lowest-order of this transformation in the spin-$1/2$ case we have 
\begin{equation}
b_{n} \rightarrow \sigma^{+}_{n} \:, \quad 
b_{n}^{\dagger} \rightarrow \sigma^{-}_{n} \:, 
\quad b_{n}^{\dagger}b_{n} \rightarrow \frac{1}{2}
\:(1-\sigma^{z}_{n}) \:.
\end{equation}
Using this last transformation, Eq.~\eqref{ExprCosine1} can be recast as 
\begin{eqnarray} \label{ExprCosine2}
\cos(\varphi_n - \varphi_{n+1})  &\approx& 1-\frac{\delta\varphi^{2}_{0}}{4}\:
\big(4-\sigma^{z}_{n}-\sigma^{z}_{n+1}\big)  \nonumber \\
&+& \frac{\delta\varphi^{2}_{0}}{2}\:\big(\sigma^{+}_{n} \sigma^{+}_{n+1}
+ \sigma^{+}_{n}\sigma^{-}_{n+1} \nonumber \\
&+& \sigma^{-}_{n}\sigma^{+}_{n+1} + \sigma^{-}_{n}\sigma^{-}_{n+1} \big) \:.
\end{eqnarray}

The terms $\sigma_{n}^{+}\sigma_{n+1}^{+}$ and $\sigma_{n}^{-}\sigma_{n+1}^{-}$ in the last expression 
can be neglected by virtue of the RWA. While this is conventionally done even in the most general (multilevel) 
treatment of transmons, in the single-fermion problem at hand such terms invariably yield zero when they
act on states in the relevant part of the Hilbert space of the system (i.e. multiqubit states with only 
one qubit in the logical $|1\rangle$ state). By also disregarding the constant term $1-\delta\varphi_{0}^{2}$
in Eq.~\eqref{ExprCosine2} which is immaterial in the present physical context, we obtain the final 
expression for $\cos(\varphi_n -\varphi_{n+1})$ [cf. Eq.~\eqref{cosform}]:
\begin{eqnarray} \label{ExprCosine3}
\cos(\varphi_{n}-\varphi_{n+1}) &\approx& \delta\varphi_{0}^{2}
\Big[\sigma_{n}^{+}\sigma_{n+1}^{-}+\sigma_{n}^{-}\sigma_{n+1}^{+} \nonumber\\
&-& \frac{\sigma_{n}^{z}+\sigma_{n+1}^{z}}{2}\Big] \:.
\end{eqnarray}

\section{Hilbert-space truncation and symmetry-adapted basis} \label{ExactDiag}
The basis of the Hilbert space $\mathcal{H}_{\text{e-ph}}=\mathcal{H}_{\text{e}}\otimes\mathcal{H}_{\text{ph}}$ 
of the coupled e-ph system under consideration is given by $|n\rangle_e\otimes|\mathbf{m}\rangle_\text{ph}$, 
with the states $|n\rangle_e\equiv c_{n}^{\dagger}|0\rangle_e$ that correspond to the excitation localized at 
the site $n$ ($n=1,\ldots,N$) and the phonon state $|\mathbf{m}\rangle_\text{ph}$ with the occupation numbers 
$\mathbf{m}\equiv(m_1,\ldots,m_N)$ at different sites:
\begin{equation}\label{mphvect}
|\mathbf{m}\rangle_\text{ph} = \prod_{n=1}^{N\otimes}
\frac{(b_n^\dagger)^{m_n}}{\sqrt{m_n!}}\:|0\rangle_\text{ph}\:.
\end{equation}
Given that the phonon Hilbert space is infinite-dimensional, we restrict ourselves to the truncated phonon 
Hilbert space. The latter includes states with the total number of phonons $m=\sum_{n=1}^N m_n$ (where $0\le m_n \le m$)
not larger than $N^{\textrm{max}}_{\textrm{ph}}$. Consequently, the dimension of the total e-ph Hilbert space 
is $D = D_\text{e} \times D_\text{ph}$, where $D_\text{e} = N$ and $D_\text{ph}=(N^{\textrm{max}}_{\textrm{ph}}+N)!
/(N^{\textrm{max}}_{\textrm{ph}}!N!)$.

The problem of diagonalizing the Hamiltonian of the coupled e-ph system under consideration can further be 
simplified by exploiting the discrete translational symmetry of this system. This symmetry is mathematically 
expressed by the commutation $[H_{\textrm{eff}},K_{\mathrm{tot}}]=0$ of the Hamiltonian $H_{\textrm{eff}}$ 
of the system and the total quasimomentum operator $K_{\mathrm{tot}}$. Owing to this symmetry, one has to 
diagonalize $H_{\textrm{eff}}$ in the sectors of $\mathcal{H}_{\text{e-ph}}$ that correspond to the eigensubspaces 
of $K_{\mathrm{tot}}$; the dimension of each of those $K$-sectors of the total Hilbert space is equal that 
of the truncated phonon space, i.e., $D_{K}=D_{\textrm{ph}}$. Accordingly, one makes use of the symmetry-adapted 
basis 
\begin{equation}\label{symmbasis}
|K,\mathbf{m}\rangle = N^{-1/2} \sum_{n=1}^N e^{iKn}\,\mathcal{T}_n(|1\rangle_\text{e} 
\otimes |\mathbf{m}\rangle_\text{ph}) \:,
\end{equation}
of $\mathcal{H}_{\text{e-ph}}$, where $\mathcal{T}_{n}$ are the discrete-translation operators;
the action of these operators has to comply with the periodic boundary conditions. The last equation
can be rewritten in the form
\begin{equation}\label{symmbasalter}
|K,\mathbf{m}\rangle = N^{-1/2} \sum_{n=1}^N e^{iKn}\, |n\rangle_\text{e} 
\otimes \mathcal{T}^{\textrm{ph}}_n|\mathbf{m}\rangle_\text{ph} \:,
\end{equation}
with the operators $\mathcal{T}^{\textrm{ph}}_n$ representing discrete translations in the phonon 
Hilbert space. If $|\mathbf{m}\rangle_\text{ph}$ is defined by a set of occupation numbers 
\begin{equation}\label{mphvectors}
|\mathbf{m}\rangle_\text{ph} = |m_1,m_2,\ldots,m_{N}\rangle_\text{ph} \:,
\end{equation}
it is straightforward to show that $\mathcal{T}^{\textrm{ph}}_n\:|\mathbf{m}\rangle_\text{ph}\equiv
|\mathcal{T}^{\textrm{ph}}_{n}\mathbf{m}\rangle$ is given by
\begin{equation}\label{mphvecttransl}
|\mathcal{T}^{\textrm{ph}}_{n}\mathbf{m}\rangle = 
|m_{N-n+1},m_{N-n+2},\ldots,m_{N-n}\rangle_\text{ph} \:.
\end{equation}

\section{Spectral-function evaluation using the KPM}   \label{KPMforSpectFunc}
\subsection{Basic aspects of the KPM} \label{KPMbasics}
In the following, we briefly recapitulate the basic aspects of the kernel polynomial method 
(KPM)~\cite{Silver+Roeder:97}, as well as the most relevant details of our concrete implementations 
thereof. A more detailed introduction into the KPM and its applications in many-body physics 
can be found in Ref.~\onlinecite{WeisseEtAlRMP:06}.

In the following, we summarize the basic aspects of the KPM along with the most relevant details 
of our own implementation of this approach for the purpose of calculating momentum-frequency 
resolved spectral function.

At the heart of the KPM lies the problem of approximating a real-valued function $f(x)$~\cite{RivlinBOOK:81}
defined on the interval $[-1,1]$ by a finite series of Chebyshev polynomials $T_n(x)$ 
of the first kind ($n = 0,\ldots,N_{\textrm{C}} - 1$):
\begin{equation}\label{ChebyshevExp}
f^{(N_{\textrm{C}})}(x)=\frac{1}{\pi\sqrt{1-x^2}}\:\left[\mu^{(0)}
+2\sum_{n=1}^{N_{\textrm{C}}-1}\mu^{(n)} T_n(x) \right] \:.
\end{equation}
The coefficients $\mu^{(n)}$ in the above expansion, referred to as Chebyshev moments, 
are given by
\begin{equation}\label{ChebyshevMoments}
\mu^{(n)}=\int_{-1}^{1} f(x)T_n(x)dx \:.
\end{equation}
For a sufficiently smooth function $f(x)$ the last series converges uniformly to 
$f$ on any closed sub-interval of $[-1,1]$ that excludes the endpoints $\pm 1$.

When the function to be approximated is not continuous or -- as is very common 
in physics applications such as the present one -- has peaks associated with 
quasiparticle states with infinite lifetime, the series in Eq.~\eqref{ChebyshevExp}
cannot converge uniformly. Namely, it fails to converge in the vicinity of a 
discontinuity, showing instead rapid oscillations whose amplitude does not decrease 
as the number of terms in the series goes to infinity (the Gibbs
phenomenon)~\cite{Shima+NakayamaBOOK:10}.

In particular, it was proven that the problem arising from the Gibbs phenomenon is solved 
for Chebyshev expansions. Namely, for any fixed number $N_{\textrm{C}}$ of terms in the 
expansion it is possible to specify a set of attenuation factors $g^{N_{\textrm{C}}}_n$ 
($n = 0,\ldots,N_{\textrm{C}}-1$), such that the modified finite-series approximants
\begin{equation}\label{ChebyshevAttenExp}
f^{(N_{\textrm{C}})}(x)=\frac{1}{\pi\sqrt{1-x^2}}\:\left[g^{N_{\textrm{C}}}_0
\mu^{(0)}+2\sum_{n=1}^{N_{\textrm{C}}-1}g^{N_{\textrm{C}}}_n \mu^{(n)} T_n(x) \right] 
\end{equation}
do not display the Gibbs phenomenon, providing instead accurate approximations 
to a broad class of functions. In other words, the introduction of the attenuation 
factors $g^{N_{\textrm{C}}}_n$ damps out high-frequency oscillations -- that would 
otherwise cause spurious results -- and constitutes the essential ingredient of the KPM.

In the problem at hand we utilize factors $g^{N_{\textrm{C}}}_n$ derived from the Jackson 
kernel~\cite{Jackson:1912}. The explicit form of those factors reads~\cite{Silver+:96}
\begin{eqnarray}
g^{N_{\textrm{C}}}_n &=& \frac{1}{N_{\textrm{C}}+1}\:\Big\{\displaystyle (N_{\textrm{C}}
-n+1)\cos\left(\frac{n\pi}{N_{\textrm{C}}+1}\right) \nonumber \\
&+& \sin\left(\frac{n\pi}{N_{\textrm{C}}+1}\right)\cot\left(\frac{\pi}{N_{\textrm{C}}+1}
\right)\Big\} \:.\label{JacksonAttenFact}
\end{eqnarray}

\subsection{Evaluation of the spectral function $A(k,\omega)$} \label{KPMbasics}
In the present work the KPM is utilized to evaluate the momentum-frequency resolved spectral 
function, given by Eq.~\eqref{CpectFuncDef}. To this end, the spectrum of the total Hamiltonian 
$H_{\textrm{eff}}=H_0+H_\text{e-ph}$ of the system ought to be mapped to the interval $[-1,1]$. 
This is accomplished by rescaling this Hamiltonian, i.e. by introducing
\begin{equation}\label{eq:H_scaled}
\tilde{H} = (1-\varepsilon)\frac{2}{\mathcal{E}_\text{max}-\mathcal{E}_\text{min}}
\left(H_{\textrm{eff}} - \frac{\mathcal{E}_\text{max}+\mathcal{E}_\text{min}}{2}\:\mathbbm{1}\right)\:,
\end{equation}
where $\mathcal{E}_\text{max}$ and $\mathcal{E}_\text{min}$ denote the largest and smallest 
eigenvalues of $H_{\textrm{eff}}$, respectively; those two eigenvalues can be obtained using the Lanczos 
algorithm~\cite{CullumWilloughbyBook}. At the same time, the parameter $\varepsilon$ is introduced 
to avoid stability problems at the boundaries of the spectrum; we hereafter set $\varepsilon=0.01$. 

By inserting the last definition of $\tilde{H}$ into the general expression for the 
spectral function [cf. Eq.~\eqref{MomFreqSpectFunc}] one arrives at 
\begin{eqnarray}\label{eq:spec_fct_scaled}
A(k,\omega) &=& \frac{2\hbar(1-\varepsilon)}
{\mathcal{E}_\text{max}-\mathcal{E}_\text{min}} \sum_j 
\left| \bra{\psi_k^{(j)}} c^\dagger_k \ket{0} \right|^2 
\notag\\&&\times\:\delta\left(\tilde{\omega}-\tilde{E}^{(j)}_k\right),
\end{eqnarray}
where $\tilde{E}^{(j)}_k$ represent the eigenvalues of $\tilde{H}$ and the rescaled 
frequency, given by
\begin{equation}\label{eq:scaled_freq}
\tilde{\omega} = \frac{2(1-\varepsilon)}{\mathcal{E}_\text{max}-\mathcal{E}_\text{min}} 
\left(\hbar\omega - \frac{\mathcal{E}_\text{max}+\mathcal{E}_\text{min}}{2}\right) \:,
\end{equation}
complies with the rescaling of the Hamiltonian in Eq.~\eqref{eq:H_scaled}.

It is worthwhile to recast Eq.~\eqref{eq:spec_fct_scaled} more succinctly as
\begin{equation}\label{eq:spec_fct_fk}
A(k,\omega) = \frac{2\hbar(1-\varepsilon)}{\mathcal{E}_\text{max}-
\mathcal{E}_\text{min}}\:f_k(\tilde{\omega}) \:,
\end{equation}
where the function $f_k(x)$ is defined as
\begin{equation}
f_k(x) \equiv \bra{0}c_k \delta(x-\tilde{H}) c^\dagger_k\ket{0} \:.
\end{equation}
The delta distribution $\delta(x-\tilde{H})$ can now be expanded in terms of Chebyshev 
polynomials $T_n(x)$ and approximated by $f^{(N_{\textrm{C}})}_k(x)$ with a large order 
$N_{\textrm{C}}$, as introduced in Eq.~\eqref{ChebyshevAttenExp}. For the sake of obtaining
a good resolution, we evaluate as many as $N_{\textrm{C}}=10^5$ moments. 

The Chebyshev moments can be derived based on the general definition in Eq.~\eqref{ChebyshevMoments}, 
which leads to the sought-after expression for the Chebyshev moments: 
\begin{equation}\label{eq:Cheb_mom_general}
\mu^{(n)}_k = \bra{0}c_k  T_n(\tilde{H}) c^\dagger_k\ket{0}  \quad 
(\:n=0,\ldots,N_{\textrm{C}}-1\:)\:.
\end{equation}
To compute these moments efficiently, it is pertinent to make use of the recurrence 
relation for Chebyshev polynomials of the first kind:~\cite{Shima+NakayamaBOOK:10}
\begin{equation} \label{Reccurence}
T_{n+1}(x) = 2x T_n(x)-T_{n-1}(x) \quad (\:n\geq 1\:)\:.
\end{equation}
Motivated by this recurrence relation, we define the states $\ket{\alpha^{(0)}_k}\equiv 
c^\dagger_k\ket{0}$ and $\ket{\alpha^{(1)}_k} \equiv \tilde{H}\ket{\alpha^{(0)}_k}$, as 
well as
\begin{equation}\label{eq:alpha_states}
\ket{\alpha^{(n+1)}_k} \equiv 2\tilde{H}\ket{\alpha^{(n)}_k} - 
\ket{\alpha^{(n-1)}_k} \:,
\end{equation}
for $n=2,\ldots,N_{\textrm{C}}-1$.

It is worthwhile noting that the states in Eq.~\eqref{eq:alpha_states} automatically occur when applying 
the recurrence relation \eqref{Reccurence} to Eq.~\eqref{eq:Cheb_mom_general}. Moreover, they render the 
computation of the Chebyshev moments rather straightforward. Namely, starting from $\mu^{(0)}_k=1$ and 
$\mu^{(1)}_k=\langle\alpha^{(1)}_k|\alpha^{(0)}_k\rangle$, one arrives at the following expressions:
\begin{eqnarray} 
\mu^{(2n)}_k &=& 2\langle\alpha^{(n)}_k |\alpha^{(n)}_k\rangle - \mu^{(0)}_k \:,\notag\\
\mu^{(2n+1)}_k &=& 2\langle\alpha^{(n+1)}_k|\alpha^{(n)}_k\rangle - \mu^{(1)}_k \:. \label{RecMoments}
\end{eqnarray}
This procedure allows a resource-friendly computation as the states $\ket{\alpha^{(n)}_k}$, being subject 
to Eq.~\eqref{eq:alpha_states}, can successively be overwritten. As a matter of fact, only three states 
$\ket{\alpha^{(n)}_k}$ have to be stored at each computational step.

It is pertinent at this point to comment on the final steps in the evaluation of the spectral function. 
In the interest of numerical efficiency, it is worthwhile to consider the following special choice of 
values for the rescaled frequency:
\begin{equation}
\tilde{\omega}_j=\cos\left[ \frac{\pi}{N_{\textrm{C}}}\left(j+\frac{1}{2}\right)\right] 
\quad (\:j=0,\ldots,N_{\textrm{C}}-1\:) \:.
\end{equation}
By inserting this last expression for $\tilde{\omega}_j$ into Eq.~\eqref{ChebyshevAttenExp} 
and making use of the identity $T_n(\cos y)=\cos(ny)$ for $y\in[-1,1]$, we obtain the following 
result for $f^{(N_{\textrm{C}})}_k(\tilde{\omega}_j)$:
\begin{eqnarray}
&&f^{(N_{\textrm{C}})}_k(\tilde{\omega}_j) =\frac{1}{\pi\sin\big[\frac{\pi}{N_{\textrm{C}}}
\big(j+\frac{1}{2}\big)\big]} \times\Big\lbrace \mu^{(0)}_k g_0 \notag\\
&+& 2\sum_{n=1}^{N_{\textrm{C}}-1}\mu^{(n)}_k g_n \cos\left[ n\frac{\pi}{N_{\textrm{C}}}
\left(j+\frac{1}{2}\right)\right]\Big\rbrace \:.
\end{eqnarray}
It is worthwhile to mention that this last expression gives rise to a discrete Fourier transformation 
that can be carried out with a modest computational effort -- more precisely, with $\mathcal{O}(N_{\textrm{C}}
\log_2 N_{\textrm{C}})$ operations -- using the well-known fast Fourier transformation (FFT) algorithm~\cite{NRcBook}.
This represents the key benefit of using Chebyshev polynomials of the first kind in the problem at hand.

Having carried out the computation of $f^{(N_{\textrm{C}})}_k(\tilde{\omega}_j)$, it remains to insert 
the obtained result into the approximated form 
\begin{equation}
A(k,\omega_j)\approx \frac{2\hbar(1-\varepsilon)}{\mathcal{E}_\text{max}-
\mathcal{E}_\text{min}} f^{(N_{\textrm{C}})}_k(\tilde{\omega}_j)\:,
\end{equation}
of Eq.~\eqref{eq:spec_fct_fk} and rescale the energy values as
\begin{equation}
\omega_j = \frac{\mathcal{E}_\text{max}-\mathcal{E}_\text{min}}{2\hbar(1-\varepsilon)}
\:\tilde{\omega}_j +\frac{\mathcal{E}_\text{max}+\mathcal{E}_\text{min}}{2\hbar} \:,
\end{equation}
which follows directly from Eq.~\eqref{eq:scaled_freq}.

\end{document}